\documentclass[a4paper,11pt]{article}
\pdfoutput=1 

\usepackage{jheppub} 
\usepackage{amsmath,amsthm,amssymb,url}

\usepackage[utf8]{inputenc}

\title{\boldmath A Gauge Field Theory for Continuous Spin Tachyons}

\author{Victor O. Rivelles}
\affiliation{ Instituto de F\'{i}sica, Universidade de S\~ao Paulo,\\ Rua do Mat\~ao, 1371, 05508-090  S\~ao Paulo, SP, Brazil}

\emailAdd{rivelles@fma.if.usp.br}

\abstract{We present a gauge field theory for the continuous spin tachyonic representation of the Poincaré group. It was  obtained by a dimensional reduction of a complex gauge field theory for a continuous spin particle in a cotangent bundle over Minkowski space-time. 
Some cubic vertices are also presented and their properties discussed. In the massless limit it 
reduces to a  gauge theory for a continuous spin particle with a global $U(1)$ symmetry. 

}

\begin{document} 

\makeatletter 
\makeatother 
		
\maketitle
\flushbottom

\section{Introduction \label{a1a} }

The irreducible unitary representations of the Poincaré group associated to particles with a finite number of physical degrees of freedom have been successfully realized as field theories in Minkowski spacetime making quantum field theory the basic tool for the study of elementary particles. However, they do not exhaust all the unitary representations of the Poincaré group 
\cite{Wigner:1939cj}. There are two classes of unitary representations with an infinite number of degrees of freedom, one being massless and named continuous (or infinite) spin particles and the other constituted by tachyonic particles. 
For a long time no field theory was known for these cases preventing the study of its properties even at the free level. Only recently a field theory for continuous spins particles was proposed \cite{Schuster:2013vpr} triggering a new wave of interest on the subject. For a recent review and earlier references see \cite{Bekaert:2017khg}.

The irreducible unitary representations of the Poincaré group can be labeled by the quadratic Casimir operator $C_2=P^2$ associated to the mass-shell condition, and the quartic Casimir operator $C_4 = -\frac{1}{2} P^2 J^{\mu\nu}J_{\mu\nu} + J^{\mu\nu} P_{\nu} J_{\mu\rho} P^{\rho}$, the square of the Pauli-Lubanski tensor \cite{Brink:2002zx,Bekaert:2006py}. If the states belonging to a given representation are labeled by $\ell$ then the unitary irreducible massless representations in four dimensions are 
\begin{center}
\begin{tabular}{l|c|c|c}
\hline
& $C_2$ & $C_4$ & $\ell$ \\
\hline
\text{helicity h} & 0 & 0 & $\pm h$ \\
\text{bosonic continuous spin} & 0 & $-\rho^2$ & 0, $\pm1, \dots \pm\infty$ \\
\text{fermionic continuous spin} & 0 & $-\rho^2$ & $\pm\frac{1}{2}, \dots \pm\infty$ \\
\hline
\end{tabular}
\end{center}
where the helicity $h$ is an integer or half-integer number and $\rho$ is a real number, the value of the continuous spin. The continuous spin particle has an infinite number of degrees of freedom and can be seen as a collection of massless fields for all helicities linked by $\rho$ since $W_\pm|0,-\rho^2,h>=\pm i \rho |0,-\rho^2;h\pm 1>$, where $W_\pm$ are the light-cone components of the Pauli-Lubanski vector. When $\rho$ vanishes the helicities become independent of each other and reduce to a set of massless particles for all helicities. It is then natural to consider a description of continuous spin particles  in terms of fields that somehow encode an infinite number of spacetime fields. The gauge theory proposed in \cite{Schuster:2013vpr} makes use of a field $\Psi(\eta,x)$ which depends not only on the spacetime coordinates $x^\mu$ but also on a extra variable $\eta^\mu$. When $\Psi(\eta,x)$ is expanded in terms of $\eta^\mu$ it naturally gives rise to an infinite number of spacetime fields. It was found that this theory is formulated in a cotangent bundle over Minkowski spacetime \cite{Rivelles:2016rwo} and that the gauge symmetries are reducible \cite{Rivelles:2014fsa}. Bosonic and fermionic continuous spin particles can also be obtained from the Fronsdal and Fang-Fronsdal equations by solving the double traceless condition \cite{Najafizadeh:2015uxa,Bekaert:2017xin}. An alternative formulation for continuous spin particles based on an oscillator formalism was presented not only in Minkowski spacetime but also extended to AdS \cite{Metsaev:2016lhs,Metsaev:2019opn,Metsaev:2021zdg}. The fermionic case \cite{Metsaev:2017ytk,Alkalaev:2018bqe,Buchbinder:2020nxn} was also considered in this framework and the supersymmetric case was considered in \cite{Buchbinder:2019kuh,Najafizadeh:2019mun,Khabarov:2020glf,Najafizadeh:2021dsm,Buchbinder:2022msd}. A frame-like formulation was presented in  \cite{Zinoviev:2017rnj,Khabarov:2017lth} while the Wigner conditions for continuous spin particles were discussed in \cite{Najafizadeh:2017tin}.  A description in terms of twistors was also proposed \cite{Buchbinder:2018soq}. Continuous spin particles with mixed symmetry, that is, not associated to totally symmetric fields, were analysed in \cite{Metsaev:2017myp,Alkalaev:2017hvj}. Cubic vertices for the interaction of continuous spin particles and massive particles, current exchange mediated by continuous spins particles and properties of its energy-momentum tensor were also investigated  \cite{Metsaev:2017cuz,Rivelles:2023hzo, Bekaert:2017xin,Rehren:2017xzn,Metsaev:2018moa,Schuster:2023xqa} while aspects of the BRST approach were discussed in \cite{Rivelles:2023hzo,Metsaev:2018lth,Buchbinder:2018yoo}.

In the massive case we also find representations similar to the continuous spin particles. We will call massive particles those with $C_2>0$ and tachyonic particles those representations which have $C_2<0$. We could disregard tachyonic particles on the basis of lack of causality. However, we must recall that they are unitary representations of the Poincaré group which can be interpreted as particles in an unstable situation that become massive particles when they reach a stable configuration. The Higgs particle is an example of such a framework. The unitary irreducible massive representations are 
\begin{center}
\begin{tabular}{ l | c | c | c  }
\hline
& $C_2$ & $C_4$ & $\ell$ \\
\hline
\text{massive spin $s$} & $m^2$ & $-m^2\,(s+1)s$ & $-s,-(s-1),\dots,s-1,s$ \\
\text{scalar tachyon} & $-m^2$ & 0 & 0  \\
\text{spin s tachyon} & $-m^2$ & $m^2\,(s+1)s$ & $\pm(s+1), \pm(s+2), \dots  \pm\infty$\\
\text{bosonic continuous spin tachyon} & $-m^2$ & $-\rho^2$ & $0,\pm 1, \pm 2, \dots \pm\infty$\\
\text{fermionic continuous spin tachyon} & $-m^2$ & $-\rho^2$ & $\pm 1/2, \pm 3/2, \dots \pm\infty$\\
\hline
\end{tabular}
\end{center} 
where $s$ is an integer or half-integer number. While the massive spin $s$ and the scalar tachyon have a finite number of degrees of freedom the remaining ones have an infinite number of them. They were discussed in the oscillator formalism \cite{Metsaev:2016lhs} but, as argued before, they should also be naturally formulated as a field theory on a cotangent bundle as we will show. 

In this paper we will present a massive gauge theory formulation for the bosonic continuous spin tachyon  along the lines of the gauge theory formulation of the continuous spin  particle of \cite{Schuster:2013vpr}. 
The tachyonic field theory will be derived in a $D+1$ cotangent bundle where 
one spacetime dimension is compactified while the corresponding coordinate in the cotangent space is left intact. The resulting action will describe a gauge theory whose structure is very similar to that of the continuous spin particle. It has two local symmetries which are reducible and the field equation allows propagation only on an hyperboloid in the cotangent bundle and its first neighborhood as it happens for the continuous spin particle. Besides, it has a global U(1) symmetry which is not required in the case  of continuous spin particles. We compute the eigenvalues of the Casimir operators to show that we are dealing with continuous spin tachyons and then make a detailed analysis of the physical degrees of freedom propagated by the field. 
Finally we discuss cubic vertices for one continuous spin tachyon and two massive scalar particles. We derive a current for the scalar fields and show that it obeys a generalized conservation condition. Solving the current conservation equation allow us to determine a local current besides a constraint among the continuous spin and the parameters that characterize the vertex. We then take the limit where the tachyon mass vanishes to get a cubic vertex for one continuous spin particle and two massive scalars recovering the results of \cite{Metsaev:2017cuz,Bekaert:2017xin}.  

The contents of the paper is as follows. In Section \ref{a2} we present a brief review of the gauge theory for continuous spin particles pointing out its main features. Then in Section \ref{p1} we present the action and local symmetries of the continuous spin tachyon showing how a convenient gauge fixing leads to simple field equations. In the next section we compute the eigenvalue of the quartic Casimir operator while in Section \ref{l5} we fix the residual local symmetries to find out the physical degrees of freedom carried by the gauge field. 
In the next section we consider the massless limit of the continuous spin tachyon and 
in the last section we discuss cubic vertices for one continuous spin tachyon and two massive scalar particles and one continuous spin particle and two massive scalar particles presentig their main properties.

\section{Continuous Spin Particles Revisited \label{a2}}

The action for a continuous spin particle is given by \cite{Schuster:2013vpr} 
\begin{equation}\label{2.1}
	S = \frac{1}{2} \int d x \, d\eta \, \delta^\prime(\eta^2+\mu^2) \left( (\partial_x \Psi(\eta,x))^2 - \frac{1}{2} (\eta^2+\mu^2) \left( \Delta \Psi(\eta,x) \right)^2 \right),
\end{equation}
where $\Delta = \partial_\eta \cdot \partial_x + \rho$ and $\delta^\prime$ is the derivative of the delta function with respect to its argument. We work in $D$ dimensions with a metric which is mostly minus. The factor $\mu^2$ was introduced to track dimensions of $\eta^\mu$ and make dimensional analysis easier. It can always be set equal to 1 by rescaling  $\eta^\mu \rightarrow \mu\eta^\mu$ and $\rho\rightarrow \rho/\mu$. The derivative of the delta function constrains the dynamics to the hyperboloid $\eta^2+\mu^2=0$ and its first neighborhood. For continuous spin particles the word hyperboloid will always refer to the $\eta^2+\mu^2=0$ hyperboloid. 

The action is invariant under the following global transformations: spacetime translations, Lorentz transformations and an $\eta^\mu$ dependent translation along $x^\mu$ given by $\delta x^\mu = \omega^{\mu\nu}\eta_\nu$, with $\omega^{\mu\nu}$ antisymmetric. This last symmetry does not preserve the natural symplectic structure of the cotangent bundle \cite{Rivelles:2016rwo}. The action is also invariant under the following local transformations
\begin{equation}\label{bbb5}
	\delta \Psi(\eta,x) = \left( \eta\cdot\partial_x - \frac{1}{2} (\eta^2+\mu^2) \Delta \right) \epsilon(\eta,x)  + \frac{1}{4} (\eta^2+\mu^2)^2 \chi(\eta,x),
\end{equation}
with $\epsilon(\eta,x)$ and $\chi(\eta,x)$ being the local parameters. All fields defined in the cotangent bundle can be expanded around the hyperboloid and the role of the $\chi$ symmetry is to remove all components of such an expansion except for the first two. It restricts the propagation of $\Psi$ to the hyperboloid and its first neighbourhood. On the other side, the $\epsilon$ symmetry is a truly gauge symmetry removing gauge degrees of freedom. These local symmetries are reducible \cite{Rivelles:2014fsa} since 
\begin{eqnarray}\label{b1}
	\delta \epsilon &=& \frac{1}{2} (\eta^2+\mu^2) \Lambda(\eta,x), \\ \label{b2}
	\delta \chi &=& \Delta \Lambda(\eta,x),
\end{eqnarray}
leave (\ref{bbb5}) invariant. This symmetry mimics the $\chi$ symmetry for $\Psi$ and can be used to limit the expansion of $\epsilon$ around the hyperboloid to just the first term.  As we shall see in the following, most of these features are shared with the tachyonic field theory. 

\section{Continuous Spin Tachyons \label{p1} }


A possible way to get a tachyonic action in $D$ dimensions is by starting with a continuous spin particle in $D+1$ dimensions. The extra spacetime coordinate $m$ is taken to be a constant and it gives rise to the tachyon mass. The extra $D+1$ $\eta$ coordinate, call it $\xi$, is left as an independent coordinate. 
By requiring the tachyonic action to have local and reducible symmetries like those similar to (\ref{bbb5}-\ref{b2}) is only possible if the field $\Psi(\eta,\xi,x)$ is complex and 
if we make the following replacements in the massless action (\ref{2.1}): $\Box_x \rightarrow \Box_x - m^2$, $ \eta^2 \rightarrow \eta^2 + \xi^2$, $\eta\cdot\partial_x \rightarrow \eta\cdot\partial_x + i m \xi$, $\Delta \rightarrow \Delta + i m\partial_\xi$ giving 
\begin{align}\label{j2}
	S = & \frac{1}{2} \int d\eta \, d\xi \, dx \,\, \delta^\prime(\eta^2+\xi^2+\mu^2) \,\, \left( \,\, |\partial_x\Psi(\eta,\xi,x)|^2 + m^2 |\Psi(\eta,\xi,x)|^2  \right. \nonumber \\
	& \left. - \frac{1}{2}(\eta^2+\xi^2+\mu^2) |(\Delta+im\partial_\xi)\Psi(\eta,\xi,x)|^2 \, \right),
\end{align} 
where $\Delta=\partial_\eta\cdot\partial_x+\rho$ as in the continuous spin case. Now the role of the derivative of the delta function is to restrict the dynamics to the hyperboloid $\eta^2+\xi^2+\mu^2=0$ and its first neighborhood. 

The action is trivially invariant under a global $U(1)$ symmetry of $\Psi$. Besides that, the action is also invariant under the local transformations
\begin{align}
	\delta\Psi(\eta,\xi,x) &= \left[ \eta\cdot\partial_x + i  m\xi - \frac{1}{2}(\eta^2+\xi^2+\mu^2) (\Delta+i m\partial_\xi)\right]\epsilon(\eta,\xi,x) \nonumber\\ &+ \frac{1}{4}(\eta^2+\xi^2+\mu^2)^2 \chi(\eta,\xi,x),\label{j3}
\end{align}
which are reducible like in the massless case
\begin{align}
	&\delta\epsilon = \frac{1}{2} (\eta^2+\xi^2+\mu^2)\Lambda, \label{j4}\\
	&\delta\chi=(\Delta+i m\partial_\xi) \Lambda. \label{j5}
\end{align}
The parameters $\epsilon(\eta,\xi,x), \chi(\eta,\xi,x)$ and $ \Lambda(\eta,\xi,x)$ are all complex. 
The $\chi$ and $\Lambda$ symmetries can be used to constrain $\Psi$ to the hyperboloid and its first neighbourhood and $\epsilon$ to the hyperboloid, respectively. The $\epsilon$ symmetry remains a gauge 
symmetry

Since $\Psi$ is complex (\ref{j2}) is of course invariant under a global $U(1)$ symmetry. We could try to remove the $U(1)$ symmetry by absorbing the $i$ factors in $m$ since they appear always multiplied by $m$. This would turn (\ref{j2}) into a complex action unless $\Psi$ is taken to be real. However, imposing a reality condition for the $\epsilon$ transformation in (\ref{j3}) would now require that the mass terms vanish leading us back to the usual continuous spin particle action and $\epsilon$ transformation. More elaborated ways to absorb the $i$'s will lead us to the same conclusion.  

The field equation for $\Psi(\eta,\xi,x)$ derived from (\ref{j2}) is
\begin{align}
	\delta^\prime(\eta^2+\xi^2+\mu^2) \left( \Box_x - m^2 - (\eta\cdot\partial_x + i m\xi)(\Delta+ i m\partial_\xi) + \frac{1}{2} (\eta^2+\xi^2+\mu^2)(\Delta+i m\partial_\xi)^2 \right) \Psi = 0. \label{j6}
\end{align}
A gauge choice which leads to a tachyonic equation for $\Psi$ is $(\Delta+i m\partial_\xi)\Psi=0$ which reduces (\ref{j6}) to $\delta^\prime(\eta^2+\xi^2+\mu^2)(\Box_x - m^2)\Psi=0$. This last equation can be solved for the delta function constraint as 
\begin{equation}
	(\Box_x-m^2)\Psi(\eta,\xi,x) - \frac{1}{4}(\eta^2+\xi^2+\mu^2)^2 \omega(\eta,\xi,x)=0, \label{j7}
\end{equation}
with $\omega(\eta,\xi,x)$ an arbitrary function. We can now use the $\chi$ symmetry to gauge $\omega$ away and this is possible only if $\chi$ satisfies $(\Box_x-m^2)\chi-\omega=0$ so that (\ref{j7}) becomes $(\Box_x-m^2)\Psi=0$. There remains a  residual $\chi_R$ symmetry with $\chi_R$ satisfying $(\Box_x-m^2)\chi_R = 0$. Besides that, the  gauge choice also imposes a further condition on $\chi_R$ so that altogether we have 
\begin{align}\label{j8}
   (\Box_x-m^2)\chi_R = 0, \\
	 \left( \eta\cdot\partial_x + i m\xi + \frac{1}{4}(\eta^2+\xi^2+\mu^2) (\Delta+i m\partial_\xi) \right)\chi_R = 0. \label{j88}
\end{align}

The gauge choice $(\Delta + i m\partial_\xi ) \Psi=0$ and $(\Box_x - m^2)\Psi=0$ impose the following constraints on $\epsilon$ 
\begin{align}
	&\left( \Box_x - m^2 - \frac{1}{2} (\eta^2+\xi^2+\mu^2)(\Delta+i m\partial_\xi)^2 \right) \epsilon=0, \label{j9}\\
	&\left( \eta\cdot\partial_x + i m\xi - \frac{1}{2} (\eta^2+\xi^2+\mu^2)(\Delta+i m\partial_\xi) \right) (\Box_x - m^2)\epsilon=0. \label{j10}
\end{align}
Having in mind the reducibility of the $\epsilon$ and $\chi$ transformations (\ref{j4}) and (\ref{j5}), we can now  make a gauge choice for $\epsilon$.  Choosing the gauge  $(\Delta+i m\partial_\xi)\epsilon=0$ (\ref{j9}) leads to $(\Box_x - m^2)\epsilon=0$, so that (\ref{j10}) is also satisfied. This gauge choice for $\epsilon$ also partially fix the $\Lambda$ symmetry so that $(\Delta+i m\partial_\xi)\epsilon=0$ and $(\Box_x - m^2)\epsilon=0$ leave a residual $\Lambda_R$ symmetry 
\begin{align}\label{j111}
	(\Box_x - m^2)\Lambda_R = 0, \\
	\left(  \eta\cdot\partial_x + i m\xi + \frac{1}{2} (\eta^2+\xi^2+\mu^2)(\Delta+i m\partial_\xi) \right) \Lambda_R = 0. \label{j11}
\end{align}
In summary we have the following partially gauge fixed set of equations and local transformations
\begin{align}
	&(\Box_x - m^2)\Psi = (\Delta+i m\partial_\xi) \Psi = 0, \label{j12}\\
	&\delta \Psi = (\eta\cdot\partial_x + i m\xi) \epsilon + \frac{1}{4}(\eta^2+\xi^2+\mu^2)^2\chi_R, \label{j13}\\
	&(\Box_x - m^2)\epsilon = (\Delta+i m\partial_\xi) \epsilon = 0, \label{j14}\\
	&\delta\epsilon = \frac{1}{2} (\eta^2+\xi^2+\mu^2)\Lambda_R, \label{j15}\\
	&\delta \chi_R = (\Delta +i m\partial_\xi ) \Lambda_R, \label{j15a}
\end{align}
where the parameters of the $\chi_R$ and $\Lambda_R$ symmetries satisfy (\ref{j8}), (\ref{j88}), (\ref{j111}) and (\ref{j11}). These residual symmetries will allow us to deal with the expansion of the fields around the hyperboloid while the $\epsilon$ symmetry will allow us to remove gauge modes. Also, some of these equations explicitly show the tachyonic nature of the fields.


To deal with the expansion of $\Psi(\eta,\xi,x)$ around the hyperboloid we will introduce new coordinates $(|\eta|,\hat{\eta}^\mu,\hat{\xi})$ defined as $\eta^\mu=|\eta|\hat{\eta}^\mu, \xi=|\eta|\hat{\xi}$ with $\hat{\eta}^\mu$ and $\hat{\xi}$ satisfying the constraint $\hat{\eta}^2+\hat{\xi}^2=-1$. Then $\hat{\eta}^\mu$ and $\hat{\xi}$ parametrizes points on the hyperboloid while $|\eta|$ parametrizes the hyperboloids. These new coordinates must be handled with care since, for instance, $\partial\hat{\eta}^\mu/\partial\hat{\eta}^\nu$ is a projection operator. 

The expansion of $\Psi(\eta,\xi,x)$ around the hyperboloid which preserves Lorentz symmetry is  
\begin{equation}\label{j17}
	\Psi(\eta,\xi,x) = \sum_{n=0}^\infty \frac{1}{n!} (\mu^2 - |\eta|^2)^n \frac{\partial^n \Psi}{\partial (\eta^2+\xi^2+\mu^2)^n}(\hat{\eta},\hat{\xi},x). 
\end{equation}
We can now expand $\chi_R$ around the hyperboloid as in (\ref{j17}) and use the $\chi_R$ symmetry in (\ref{j13}) to gauge away all terms with $n\ge 2$ in the expansion of $\Psi$ (\ref{j17}) yielding 
\begin{equation}\label{j18}
	\Psi_{\chi}(\eta,\xi,x) = \Psi(\hat{\eta},\hat{\xi},x) + (\mu^2- |\eta|^2)  \frac{\partial \Psi}{\partial (\eta^2+\xi^2+\mu^2)}(\hat{\eta},\hat{\xi},x),
\end{equation}
where $\Psi_\chi(\eta,\xi,x)$ means the $\chi_R$ fixed form of $\Psi(\eta,\xi,x)$. This procedure has to be compatible with both equations in (\ref{j12}) and this happens if $\chi_R$ satisfies (\ref{j8}) and (\ref{j88}). Since all terms in the expansion of $\chi_R$ were used the $\chi_R$ symmetry is completely fixed. We then find that the first equation in (\ref{j12}) leads to 
\begin{align}\label{j19}
	(\Box_x-m^2) \Psi(\hat{\eta},\hat{\xi},x) = 0, \\
	(\Box_x-m^2) \frac{\partial \Psi}{\partial(\eta^2+\xi^2+\mu^2)}(\hat{\eta},\hat{\xi},x) = 0, \label{j199}
\end{align}
showing that $\Psi(\eta,\xi,x)$ propagates only on the hyperboloid and its first neighbourhood. 

The same procedure can be applied to $\epsilon$ to show that when expanded around the hyperboloid only the first term survives 
\begin{align}
	\epsilon_\Lambda(\eta,\xi,x) = \epsilon(\hat{\eta},\hat{\xi},x), 
\end{align}
with $\epsilon_\Lambda(\eta,\xi,x)$ meaning the $\Lambda_R$ gauge fixed form of $\epsilon(\eta,\xi,x)$. As for the $\chi_R$ symmetry, we have to show that this procedure is consistent with both equations in (\ref{j14})  and this is true if $\Lambda_R$ satisfies (\ref{j111}) and (\ref{j11}). Then the $\Lambda_R$ symmetry is also completely fixed. The gauge symmetry in (\ref{j13}) then becomes 
\begin{align}
	&\delta \Psi(\hat{\eta},\hat{\xi},x) = \mu (\hat{\eta}\cdot\partial_x + i m \hat{\xi})\epsilon(\hat{\eta},\hat{\xi},x), \label{b5a} \\
	&\delta \frac{\partial \Psi}{\partial (\eta^2+\hat{\xi}^2+\mu^2)}(\hat{\eta},\hat{\xi},x) = - \frac{1}{2\mu} (\hat{\eta}\cdot\partial_x + i m\hat{\xi}) \epsilon(\hat{\eta},\hat{\xi},x). \label{b6a}
\end{align}

\section{Casimir Operators \label{j20}}

In order to find out which representations of the Poincaré group are being carried by $\Psi$ we have to compute the Casimir operators. The relevant ones are the quadratic $C_2$ and the quartic $C_4$ Casimir operators associated to the mass and spin contents of $\Psi$ respectively. 

The quadratic Casimir is $C_2=P^2=-\Box_x $ and using the field equation (\ref{j6}) we get
\begin{align}
	\delta^\prime(\eta^2+\xi^2+\mu^2) C_2 \Psi = \delta^\prime(\eta^2+\xi^2+\mu^2) \left(- m^2\Psi + \delta_\epsilon\Psi \right), \label{212} 
\end{align}
with $\epsilon = - (\Delta + i m \partial_\xi )\Psi$,
so that we are dealing with a tachyonic representation. 

The quartic Casimir operator receives no contribution from the extra coordinate $\xi$ since  
$J_{\mu\nu}= i x_{[\mu} \partial_{x\mu]} + i \eta_{[\mu} \partial_{\eta\nu]}$ so we get  
\begin{align}\label{21}
	 C_4 = &\left[ (D-3+\eta\cdot\partial_\eta)\eta\cdot\partial_\eta - \eta^2\Box_\eta)\right]\Box_x 
	-\eta\cdot\partial_x (D-2+2\eta\cdot\partial_\eta)\partial_\eta\cdot\partial_x \nonumber \\
	&+ (\eta\cdot\partial_x)^2 \Box_\eta + \eta^2 (\partial_\eta\cdot\partial_x)^2.
\end{align}
Using again (\ref{j6}) we find that
\begin{equation} \label{22}
	\delta^\prime(\eta^2+\xi^2+\mu^2) C_4 \Psi = \delta^\prime(\eta^2+\xi^2+\mu^2) \left( -\mu^2\rho^2 \Psi + \delta_\epsilon \Psi  \right),
\end{equation}
 with
\begin{align}\label{23}
\epsilon  =  & (D-3+\eta\cdot\partial_\eta) \eta\cdot\partial_\eta (\Delta - im\partial_\xi) + (D-2+2\eta\cdot\partial_\eta) (\rho - im\partial_\xi) \nonumber \\
& + [ (\mu^2+\xi^2) (\Delta - im\partial_\xi) + (\eta\cdot\partial_x + im\xi) ]\Box_\eta,
\end{align}
Then (\ref{22}) shows that $\Psi$ has the quartic Casimir of a bosonic continuous spin tachyon. 

Finally, taking into account the expansion of $\Psi$ around the hyperboloid (\ref{j18}) we find that the quartic Casimir operator acts on the hyperboloid and its first neighbourhood as 
\begin{align}
	&C_4 \Psi(\hat{\eta},\hat{\xi},x) = -\mu^2\rho^2 \Psi(\hat{\eta},\hat{\xi},x), \label{d5a}\\
	&C_4 \frac{\partial \Psi}{\partial(\eta^2+\xi^2+\mu^2)} (\hat{\eta},\hat{\xi},x) = -\mu^2\rho^2 \frac{\partial \Psi}{\partial(\eta^2+\xi^2+\mu^2)} (\hat{\eta},\hat{\xi},x), \label{d6a}
\end{align}
up to $\epsilon$ gauge transformations.

\section{Physical Degrees of Freedom \label{l5}}

In order to find the physical degrees of freedom carried by $\Psi(\eta,\xi,x)$ we have to solve (\ref{j12}) and (\ref{j14}). Let us do this before fixing the residual local symmetries $\chi_R$ and $\Lambda_R$. In the rest frame the first equation in (\ref{j12}) can be solved in momentum space as $k_{D-1}=m, k_\alpha =0, \alpha=0, \dots D-2$. The $\epsilon$ gauge transformation in (\ref{j13}) shows that the first equation of (\ref{j14}) must be solved for the same momentum. 

To solve the second equation in (\ref{j12}) it is better to change the $\eta_{D-1}$ and $\xi$ coordinates to $\xi_\pm = (\xi \pm \eta_{D-1})/2$ so that the solution is 
\begin{equation}
	\Psi(\eta,\xi,k) = e^{\frac{i\rho}{m}\xi_+} \psi(\eta_\alpha,\xi_-,k_{D-1}). \label{k4}
\end{equation}
In a similar way we solve the second equation in (\ref{j14}) as
\begin{equation}
	\epsilon(\eta,\xi,k) = e^{\frac{i\rho}{m}\xi_+} \varepsilon(\eta_\alpha,\xi_-,k_{D-1}). \label{k5}
\end{equation}
Then the $\epsilon$ gauge transformation (\ref{j13}) turns into 
\begin{equation}
	\delta\psi(\eta_\alpha,\xi_-,k_{D-1}) =  2im\xi_- \varepsilon(\eta_\alpha,\xi_-,k_{D-1}), \label{k6}
\end{equation}
so that all $\xi_-$ dependence of $\psi(\eta_\alpha,\xi_-,k_{D-1})$ can be gauged away and we have 
\begin{equation}
	\Psi(\eta,\xi,k) = e^{\frac{i\rho}{m}\xi_+} \psi(\eta_\alpha,k_{D-1}). \label{k7}
\end{equation}

We can now use the $\chi_R$ and $\Lambda_R$ symmetries to restrict the dynamics to the hyperboloid. Expanding $\Psi$ as in (\ref{j18}) we find that 
\begin{align}
	\Psi(\hat{\eta},\hat{\xi},k) &= e^{\frac{i\rho\mu}{m}\hat{\xi}_+} \psi(\hat{\eta}_\alpha,k_{D-1}), \label{l1} \\
	\frac{\partial\Psi}{\partial(\eta^2+\xi^2+\mu^2)}(\hat{\eta},\hat{\xi},k) &= 
	-\frac{1}{2\mu} e^{\frac{i\rho\mu}{m}\hat{\xi}_+} \left( \frac{i\rho}{m} \psi(\hat{\eta}_\alpha,k_{D-1}) + \frac{\partial\psi}{\partial|\eta|}(\hat{\eta}_\alpha,k_{D-1}) \right). \label{l11}
\end{align}
At this point it is important to analyze the dependence of these expressions with respect to $\hat{\eta}_\alpha$. Let us assume that $\psi(\hat{\eta}_\alpha,k_{D-1})$ and $\frac{\partial\psi}{\partial{|\eta |}}(\hat{\eta}_\alpha,k_{D-1})$ do depend on $\hat{\eta}^\alpha \hat{\eta}_\alpha$. Then from the constraint $\hat{\eta}^2 + \hat{\xi}^2=-1$ we find that $\hat{\eta}^\alpha \hat{\eta}_\alpha = -1-\hat{\xi}_+\hat{\xi}_-$ so that we can use (\ref{k6}) to gauge away all $\hat{\xi}_+$ dependence. This means that $\psi(\hat{\eta}_\alpha,k_{D-1})$ and $\frac{\partial\psi}{\partial{|\eta |}}(\hat{\eta}_\alpha,k_{D-1})$ do not depend on $\hat{\eta}^\alpha \hat{\eta}_\alpha$ since effectively $\hat{\eta}^\alpha \hat{\eta}_\alpha$ reduces to $-1$. As a consequence, the expansion of $\psi(\hat{\eta}_\alpha,k_{D-1})$ and $\frac{\partial\psi}{\partial{|\eta |}}(\hat{\eta}_\alpha,k_{D-1})$ in powers of  $\hat{\eta}_\alpha$ do not generate trace contributions which implies that the polarization tensors on the hyperboloid and its first neighborhood are symmetric and traceless. We then find that 
\begin{align}\label{l6}
	\Psi_\chi(\eta,\xi,k) = e^{ \frac{i \rho\mu\hat{\xi}_+}{m}} \sum_{n=0}^\infty \frac{\mu^n}{n!} \hat{\eta}^\alpha_1 \cdots \hat{\eta}^\alpha_n \left[ 1  - \frac{1}{2\mu}(\mu^2-|\eta|^2) \left( \frac{n}{\mu} + \frac{i \rho}{m}  	\right)	\right] \psi^T_{\alpha_1\cdots\alpha_n}(k_{D-1}),
\end{align}
with $\psi^T_{\alpha_1\cdots\alpha_n}(k_{D-1})$ traceless.

Summarizing, we have found that after solving the field equation and the gauge fixing condition (\ref{j12}) $\Psi(\eta,\xi,k)$ can be written in terms of $\psi(\eta_\alpha,k_{D-1})$ as in (\ref{k7}). The components of $\psi(\eta_\alpha,k_{D-1})$, that is $\psi_{\alpha_1\dots\alpha_n}(k_{D-1})$, are not traceless so that all helicities are present and each one appears an infinite number of times. This happens because the $\chi$ symmetry was not fixed yet. Fixing the $\chi$ symmetry leads to the  expansion around the hyperboloid (\ref{j18}) where $\Psi(\eta,\xi,x)$ and its derivative must be written in terms of $\psi(\hat{\eta}_\alpha,k_{D-1})$ as in (\ref{l1}) and (\ref{l11}). We can now write $\psi(\hat{\eta}_{\alpha},k_{D-1})$ and its first derivative in terms of its components $\psi_{\alpha_1\dots\alpha_n}(k_{D-1})$ to find (\ref{l6}). This equation shows that $\Psi_\chi$, with the $\chi$ and $\epsilon$ symmetries fixed, has only traceless components, that is $\psi^T_{\alpha_1\dots\alpha_n}(k_{D-1})$, so that we have the expected number of degrees of freedom for a continuous spin tachyon. Also, it shows how they are distributed on the hyperboloid and its first neighborhood. On the hyperboloid we have a particular linear combination of the traceless components of $\psi(\hat{\eta}_\alpha,k_{D-1})$. At the first neighborhood we have another linear combination of components of $\psi(\hat{\eta}_\alpha,k_{D-1})$ since the coefficients of the linear combination 
depends on $n$. Hence the same degrees of freedom propagate on the hyperboloid and its first neighborhood but they appear in different linear combinations. Notice that this result is gauge dependent. Other gauge choices will change the linear combinations appearing on the hyperboloid and its first neighborhood.  
%
%
Since $\Psi(\eta,\xi,x)$ is complex it describes a $U(1)$ charged continuous spin tachyon with continuous spin $\rho$ propagating on the hyperboloid and its first neighborhood. 

For $\rho\rightarrow 0$ (\ref{22}) shows that we have a tachyonic representation with an infinite number of degrees of freedom with $C_4=0$. This is a non-unitary representation since the only unitary tachyonic representation with $C_4=0$ is the scalar tachyon which carries one degree of freedom. Therefore, the $\rho\rightarrow 0$ limit produces a field theory which does not propagate any physical degree of freedom.

We could wonder whether it is possible to describe spin $s$ tachyons in this set up. It is worth recalling that in the oscillator formalism \cite{Metsaev:2016lhs} spin $s$ tachyons always appear together with non-unitary representations. Here, the first step is to make sure that the quartic Casimir has the right expression for a spin s tachyon. This is possible only if $\rho^2 = - \frac{m^2}{\mu^2}(D-3+s)s$ which means that $\rho$ is pure imaginary and that is not acceptable. If we proceed like in  \cite{Metsaev:2016lhs}, the sum in (\ref{l6}) can be split as 
%
%
\begin{align}\label{l8}
&\Psi_\chi(\eta,\xi,k) = \nonumber \\
&e^{\frac{i \rho\mu\hat{\xi}_+}{m}} \left(\sum_{n=0}^s + \sum_{n=s+1}^\infty \right)   \frac{\mu^n}{n!} \hat{\eta}^\alpha_1 \cdots \hat{\eta}^\alpha_n \left[ 1  - \frac{1}{2\mu}(\mu^2-|\eta|^2) \left( \frac{n}{\mu} + \frac{i \rho}{m}  	\right)	\right] \psi^T_{\alpha_1\cdots\alpha_n}(k_{D-1}).
\end{align}
If ${s=0}$ then $C_4=0$ and the first sum has only one term which is describing a scalar tachyon.  The second sum is a non-unitary representation because it has an infinite number of degrees of freedom and has $C_4=0$. If ${s>0}$ we find that in the second sum the propagating degrees of freedom have spins from $s+1$ to $\infty$ so it could describe a spin $s$ tachyon. The first sum, however, has a finite number of degrees of freedom so it is a non-unitary tachyonic representation. Therefore, it is not possible to describe a spin $s$ tachyonic representation with the present formalism.  


We could also try to describe massive particles. To consider massive representations we just have to replace $m^2 \rightarrow - m^2$, $\xi^2 \rightarrow - \xi^2$, $m\partial_{\xi} \rightarrow -m\partial_{\xi}$ and $\xi_\pm \rightarrow \pm \xi_\mp$ in all previous equations. The counting of physical degrees of freedom is done in a similar way with the $SO(1,D-2)$ representations replaced by $SO(D-1)$ representations. 
%
%
For $\rho\not= 0$, $\Psi$ is describing a massive representation with infinite degrees of freedom so it is non-unitary. For $\rho\rightarrow 0$ we have $C_4=0$ and we would expect to find  a single massive spin $0$ representation. However $\Psi$ has components with spins ranging from $-\infty$ to $+\infty$ with each spin appearing just once. They do not form a sum of massive representations for all spins so it is also a non-unitary representation. 
%
If $\rho^2 = \frac{m^2}{\mu^2}(D-3+s)s$ we also find a reducible representation if we split the sum as in (\ref{l8}). The first sum describes a spin $s$ massive particle while the second sum, being an infinite sum, is describing a non-unitary massive representation. Then in the massive case $\Psi$ always carries non unitary representations.

\section{Massless limit  \label{ml}}


When taking the massless limit, that is, $m\rightarrow 0$, $\xi$ $\rightarrow 0$, and also considering $\Psi(\eta,x)$ real, we expect that all previous equations reproduce known results for the continuous spin particles. Since the gauge choice used here is much better than those discussed 
in previous papers on continuous spin particles we will make a brief presentation for the massless case calling attention to the new points. 

The field equation obtained from (\ref{2.1}) is 
\begin{equation}
	\delta^\prime(\eta^2+\mu^2) \left( \Box_x  - \eta\cdot\partial_x \Delta + \frac{1}{2} (\eta^2+\mu^2) \Delta^2 \right)\Psi = 0, \label{i3}
\end{equation}
and the harmonic gauge choice $\Delta \Psi=0$ reduces it to $\delta^\prime(\eta^2+\mu^2) \Box_x \Psi = 0$. The delta function constraint can be solved as
\begin{equation} \label{a3}
	\Box_x \Psi(\eta,x) - \frac{1}{4} (\eta^2+\mu^2)^2 \omega(\eta,x)=0,
\end{equation}
where $\omega(\eta,x)$ is an arbitrary function. We can now use the $\chi$ symmetry of (\ref{bbb5}) to remove $\omega(\eta,x)$ but leaving a residual $\chi_R$ symmetry satisfying $\Box_x \chi_R=0$. Consistency  with the harmonic gauge choice then requires 
\begin{equation}\label{a4}
	 \Box_x \chi_R = (\eta^2+\mu^2) \left(\eta\cdot\partial_x + \frac{1}{4} (\eta^2+\mu^2)\Delta \right)\chi_R = 0.
\end{equation}

The $\epsilon$ gauge symmetry is consistent with the harmonic gauge choice and with $\Box_x \Psi=0$ if
\begin{equation}\label{a5}
	\left(\eta\cdot\partial_x - \frac{1}{2} (\eta^2+\mu^2)\Delta \right) \Box_x\epsilon = \left(\Box_x - \frac{1}{4} (\eta^2+\mu^2)\Delta^2 \right) \epsilon= 0,
\end{equation}
respectively. The reducibility of the $\epsilon$ and $\chi$ transformations (\ref{b1}-\ref{b2}) allow us to choose an harmonic gauge for $\epsilon$, $\Delta\epsilon=0$, which leads to $\Box_x\epsilon=0$. The first equation in (\ref{a5}) is then also satisfied. These equations partially fix the $\Lambda$ symmetry in (\ref{b1}-\ref{b2}) leaving a residual $\Lambda_R$ symmetry satisfying 
\begin{equation}\label{a6}
	(\eta^2+\mu^2) \Box_x \Lambda_R = \left( \eta\cdot\partial_x - \frac{1}{2} (\eta^2+\mu^2)\Delta \right) \Lambda_R = 0.
\end{equation}

To summarize, the harmonic gauge choice for the continuous spin particle leads to
\begin{eqnarray}
	&&\Box_x \Psi = \Delta \Psi = 0, \label{a7} \\
	&&\delta \Psi = \eta\cdot\partial_x \epsilon + \frac{1}{4} (\eta^2+\mu^2)^2 \chi_R, \label{a9}\\
  &&\Box_x \epsilon = \Delta \epsilon = 0, \label{a8} \\
	&&\delta \epsilon = \frac{1}{2} (\eta^2+\mu^2) \Lambda_R(\eta,x), \label{g2} \\
	&& \delta \chi_R = \Delta \Lambda_R, \label{g2a}
\end{eqnarray}
with the residual $\chi_R$ and $\Lambda_R$ transformation parameters satisfying (\ref{a4}) and (\ref{a6}) respectively. As in the tachyonic case these residual transformations will allow us to deal with the expansion of the fields around the hyperboloid $\eta^2+\mu^2=0$.

The expansion around the hyperboloid requires the use of coordinates $(|\eta|, \hat{\eta}^\mu)$ defined by $\eta^\mu = |\eta| \hat{\eta}^\mu$ with $\hat{\eta}^2 =-1$. Then $\hat{\eta}^\mu$ parametrizes points on the hyperboloid while $|\eta|$ parametrizes different hyperboloids. The expansion of $\Psi(\eta,x)$ around the hyperboloid is then 
\begin{equation}\label{b3}
	\Psi(\eta,x) = \sum_{n=0}^{\infty} \frac{1}{n!} (\mu^2-|\eta|^2)^n \frac{\partial^n \Psi}{\partial (\eta^2+\mu^2)^n}(\hat{\eta},x).
\end{equation}
We can now expand $\chi_R$ around the hyperboloid as in (\ref{b3}) and use the $\chi_R$ transformation of (\ref{a9}) to gauge away all terms in (\ref{b3}) with $n\geq 2$ so that  
\begin{equation}\label{b7}
	\Psi_{\chi}(\eta,x) = \Psi(\hat{\eta},x) + (\mu^2-|\eta|^2)  \frac{\partial \Psi}{\partial (\eta^2+\mu^2)}(\hat{\eta},x),
\end{equation}
where $\Psi_{\chi}(\eta,x)$ stands for the $\chi_R$ gauge fixed form of $\Psi(\eta,x)$. 
We have to check the compatibility of this procedure with both equations in (\ref{a7}) and this happens if $\chi_R$ satisfies both equations in (\ref{a4}). Then the $\chi_R$ symmetry is completely fixed. The first equation in (\ref{a7}) then leads to
\begin{equation}\label{b4}
	\Box_x \Psi(\hat{\eta},x) = \Box_x \frac{\partial \Psi}{\partial(\eta^2+\mu^2)}(\hat{\eta},x) = 0,
\end{equation}
showing that $\Psi$ propagates on the hyperboloid and its first neighbourhood.

As in the tachyonic case, the same procedure can be applied to $\epsilon$ and $\Lambda_R$ so that 
\begin{equation}\label{b8}
	\epsilon_{\Lambda}(\eta,x) = \epsilon(\hat{\eta},x), 
\end{equation}
where $\epsilon_{\Lambda}(\eta,x)$ means the $\Lambda_R$ gauge fixed form of $\epsilon(\eta,x)$. As for the $\chi_R$ symmetry, we have to show that this procedure is consistent with both equations in (\ref{a6}) and this will happen if both equations for $\Lambda_R$ in (\ref{a5}) are satisfied. The $\Lambda_R$ symmetry is  also completely fixed. Finally, the gauge symmetry in (\ref{a9}) becomes
\begin{align}
	&\delta \Psi(\hat{\eta},x) = \mu \hat{\eta}\cdot\partial_x \epsilon(\hat{\eta},x), \label{b5} \\
	&\delta \frac{\partial \Psi}{\partial (\eta^2+\mu^2)}(\hat{\eta},x) = - \frac{1}{2\mu} \hat{\eta}\cdot\partial_x \epsilon(\hat{\eta},x). \label{b6}
\end{align}

The quartic Casimir operator is now 
\begin{equation}\label{d2}
  C_4 = (D-3+\eta\cdot\partial_\eta)\eta\cdot\partial_\eta \Box_x - \eta^2 \Box_\eta\Box_x - \eta\cdot\partial_x (D-2+2\eta\cdot\eta) \partial_\eta\cdot\partial_x + (\eta\cdot\partial_x)^2 \Box_\eta + \eta^2 (\partial_\eta\cdot\partial_x)^2,
\end{equation}
and making use of (\ref{a7}) we find 
\begin{equation} \label{d3}
	C_4 \Psi(\eta,x) = -\mu^2\rho^2 \Psi(\eta,x) + \delta_\epsilon \Psi(\eta,x), 
\end{equation}
where the parameter for the gauge transformation is 
\begin{equation}\label{d4}
	\epsilon(\eta,x) = \left[ \rho(D-2+2\eta\cdot\partial_\eta) + \eta\cdot\partial_x \Box_\eta\right]\Psi(\eta,x). 
\end{equation}
We can now use (\ref{b7}) to find out how the Casimir operator acts on the hyperboloid and its first neighbourhood obtaining 
\begin{align}
	&C_4 \Psi(\hat{\eta},x) = -\mu^2\rho^2 \Psi(\hat{\eta},x), \label{d5}\\
	&C_4 \frac{\partial \Psi}{\partial(\eta^2+\mu^2)} (\hat{\eta},x) = -\mu^2\rho^2 \frac{\partial \Psi}{\partial(\eta^2+\mu^2)} (\hat{\eta},x), \label{d6}
\end{align}
up to $\epsilon$ gauge transformations.

We can now go back to (\ref{a7})-(\ref{g2a}) where the residual $\chi_R$ and $\Lambda_R$ symmetries were not fixed yet. We can solve the first equation in (\ref{a7}) by choosing a light-like momentum  with light-cone components $k_+\not= 0, k_-=k_i=0, i=1,\dots D-2$. Then the $\epsilon$ gauge symmetry in (\ref{a9}) tell us that the solution of the first equation in (\ref{a8}) implies that $\epsilon$ must have the same momentum as $\Psi$ and that in momentum space 
\begin{equation}\label{h1}
	\delta \Psi(\eta,k_+) = i k_+ \eta_- \epsilon(\eta,k_+).
\end{equation}
We can then solve the second equation in (\ref{a7}) and (\ref{a8}) as
\begin{eqnarray}
	&&\Psi(\eta,k_+) = e^{-\frac{\rho\eta_+}{ik_+}} \psi(\eta_-,\eta_i,k_+), \label{h2}\\
	&&\epsilon(\eta,k_+) = e^{-\frac{\rho\eta_+}{ik_+}} \varepsilon(\eta_-,\eta_i,k_+), \label{h3}
\end{eqnarray}
so that (\ref{h1}) now reads
\begin{equation}\label{i1}
	\delta \psi(\eta_-,\eta_i,k_+) = i k_+ \eta_- \varepsilon(\eta_-,\eta_i,k_+).
\end{equation}
The gauge transformation (\ref{i1}) shows that all terms proportional to $\eta_-$ in $\psi(\eta_-,\eta_i,k_+)$ can be gauged away so that it depends only on $\eta_i$ 
\begin{equation}\label{h4}
	\Psi(\eta,k_+) = e^{-\frac{\rho\eta_+}{ik_+}} \psi(\eta_i,k_+). 
\end{equation}

We can now use the expansion of $\Psi(\eta,x)$ in (\ref{b7}) to find that on the hyperboloid and its first neighbourhood we have
\begin{align}
	\Psi_{\chi}({\eta},k) = e^{-\frac{\rho\mu\hat{\eta}_+}{ik_+}} \left[ \psi(\hat{\eta}_i,k_+) -\frac{1}{2\mu} (\mu^2-|\eta|^2)  \left( - \frac{\rho\hat{\eta}_+}{ik_+} \psi(\hat{\eta}_i,k_+) + \frac{\partial\psi}{\partial{|\eta |}}(\hat{\eta}_i,k_+) \right) \right] . \label{ii3}
\end{align}
Like in the massive case we can show that  $\psi(\hat{\eta}_i,k_+)$ and $\frac{\partial\psi}{\partial{|\eta |}}(\hat{\eta}_i,k_+)$ does not depend on $(\hat{\eta}_i)^2$ and hence their expansion in powers of $\hat{\eta}_i$ has only symmetric traceless tensors. The argument runs along the same lines as for the massive case but now we must use that  $(\hat{\eta}_i)^2= 1+2\hat{\eta}_+\hat{\eta}_-$ and consider that the gauge symmetry (\ref{i1}) allow us to gauge away all terms in $\hat{\eta}_-$. We then have 
\begin{align}\label{k1}
	\Psi_{\chi}({\eta},k) = e^{-\frac{\rho\mu\hat{\eta}_+}{ik_+}} \sum_{n=0}^\infty \frac{\mu^n}{n!} \hat{\eta}^{i_1}\cdots\hat{\eta}^{i_n} \left[ 1 - \frac{1}{2\mu} (\mu^2-|\eta|^2) \left( \frac{n}{\mu} - \frac{\rho\hat{\eta}_+}{ik_+} ) \right)  \right] \psi^T_{i_1\dots i_n}(k_+).
\end{align}

Now the polarization tensors form finite dimensional representations of $SO(D-2)$ so that in 4 dimensions they carry integer helicities from $-\infty$ to $\infty$ which is the expected contents of a continuous spin particle for $\rho\not= 0$. 
Then $\Psi(\eta,x)$ describes a single continuous spin particle propagating on the hyperboloid and its first neighbourhood. 

If we take the limit $\rho\rightarrow 0$, which means from (\ref{d3}) that we have a reducible representation for massless particles, we find the polarizations tensors for massless fields with integer helicities from $-\infty$ to $\infty$. Notice however that this produces a very peculiar set of massless higher spin particles since all of them have the same momentum and propagate in the same direction. This happens because the fundamental field $\Psi(\eta,x)$ satisfies $\Box_x\Psi(\eta,x)=0$ so that all polarizations in (\ref{k1}) have the same momentum. On the other side if we go back to the action (\ref{2.1}) and start with $\rho=0$ from the beginning we can show that the field equation reduces to a sum of Fronsdal actions \cite{Rivelles:2014fsa}. In this situation the spacetime fields are independent of each other since each field has its own momentum. Back to the continuous spin case we see that this independence does not hold because for $\rho\not= 0$ the  solution of the second equation in (\ref{a7}) relates polarization tensors of different orders, that is, it express  the {\it minus} components of the polarization tensor to its $+$ and $i$ components \footnote{Recall that the $+$ components are gauged away by (\ref{i1}).}.

\section{Cubic Vertices \label{s1}}


In this section we will discuss cubic vertices involving continuous spins tachyons and massive scalar particles. We will not try to present a systematic analysis but rather just to point out its main features providing some simple examples. We will start with continuous spin tachyons and then take the massless limit to get vertices for continuous spin particles. Since continuous spin tachyons and continuous spin particles are described by gauge theories they give rise to conserved currents which will be used to restrict the form of the vertices. Also we will consider only parity invariant terms in the cubic action.

Let us consider a cubic vertex for the continuous spin tachyon 
\begin{align}
 S_c = g \int d\eta\,d\xi\,dx \,\, \delta^{\prime}(\eta^2+\xi^2+\mu^2) \, \Psi(\eta,\xi,x) \, J(\eta,\xi,x) + c.c., \label{s3}
\end{align}
where $J(\eta,\xi,x)$ depends on two complex massive scalar fields and $g$ is a coupling constant. This vertex is invariant under the $\chi$ symmetry of $\Psi$ (\ref{j3}) and a $\chi$ type symmetry of $J$
\begin{align}
	\delta J(\eta,\xi,x) = \frac{1}{4} (\eta^2+\xi^2+\mu^2)^2 \Xi(\eta,\xi,x), \label{v1}
\end{align}
where $\Xi(\eta,\xi,x)$ is a complex arbitrary function. This symmetry is not due to a $\chi$ type symmetry for the scalar fields but to the delta function structure of the vertex only. Notice that if we had higher derivatives of the delta function in (\ref{s3}), say $p>1$ derivatives, we could multiply the field equation by $(\eta^2+\xi^2+\mu^2)^p$ and find that $J$ vanishes on the hyperboloid. Only for $p=0,1$ we have a non vanishing $J$ so that we took the highest allowed value of $p$ in (\ref{s3}). Also, only for $p=0,1$ the vertex is invariant under the $\chi$ symmetry of $\Psi$. 

The continuous spin tachyon field equation is now 
\begin{align}
	&\delta^\prime(\eta^2+\xi^2+\mu^2) \left[ \left(\Box_x - m^2 - (\eta\cdot\partial_x + i m\xi)(\Delta+i m \partial_\xi) + \frac{1}{2} (\eta^2+\xi^2+\mu^2)(\Delta+ i m\partial_\xi)^2 \right)\Psi \right.\nonumber \\
	& \left. - g\,J \right] = 0. \label{s5}
\end{align}
We can multiply it by $\eta^2+\xi^2+\mu^2$ and apply $\Delta+i m\partial_\xi$ to get 
\begin{align}
	\delta(\eta^2+\xi^2+\mu^2) (\Delta+i m\partial_\xi) J(\eta,\xi,x)=0, \label{s6}
\end{align}
which is the generalization of a current conservation equation to the cotangent bundle. Notice also that the cubic vertex (\ref{s3}) is invariant under the $\epsilon$ gauge transformation (\ref{j3}) if (\ref{s6}) holds. 
Taking (\ref{v1}) into account we can expand $J(\eta,\xi,x)$ as 
\begin{align}
J(\eta,\xi,x) = J_0(\eta,\xi,x) +  (\eta^2+\xi^2+\mu^2) J_1(\eta,\xi,x), \label{w1}
\end{align}
and use (\ref{s6}) to get 
\begin{align}
	\delta(\eta^2+\xi^2+\mu^2) \left[ (\Delta+i m\partial_\xi) J_1(\eta,\xi,x) + 2(\eta\cdot\partial_x + i m\xi) J_0(\eta,\xi,x) \right] = 0. \label{w2}
\end{align}
Now we have to find two currents $J_0$ and $J_1$ depending on two complex scalar fields satisfying the conservation condition (\ref{w2}). 
The field equations for the complex scalar fields $\phi_i(x), i=1,2$, with mass $M_i$ are 
\begin{align}
	(\Box_x + M_i^2) \phi_{i}(x) =  {\cal O}(g), \label{s7}
\end{align}
where the explicit form of the ${\cal O}(g)$ terms are not needed. 

The next step is to find a solution for the currents in (\ref{w2}) using the field equations (\ref{s5}) and (\ref{s7}). We will proceed by proposing an ansatz for $J_1$ and use (\ref{w2}) to determine $J_0$. The solution will also relate $\rho$ to the parameters in $J_0$ and $J_1$. 

The currents $J_0$ and $J_1$ must depend on the scalar fields and its derivatives. Lorentz invariance thus requires that we use the operators $\Box_x$ and $\eta\cdot\partial_x+im\xi$ which are present in (\ref{w2}).  Since $\Box_x$ acting on the scalar field is proportional to the field itself the derivatives must act on different fields so that the simplest situation is that in which $J_0$ depends only on $\eta\cdot\partial_x+im\xi$. Notice also that in (\ref{w2}) we have derivatives of $J_0$ and $J_1$ and to have a chance that they cancel out we should use exponentials of $\eta\cdot\partial_x+im\xi$ wherever it is possible. Taking all this into account the simplest ansatz for $J_0$ is then
\begin{align}
	J_0(\eta,\xi,x) = (\eta\cdot\partial_x+im\xi)^{n_0} \left( f_{\lambda_1}^{n_1} \phi_{1}(x) \,\,\, f_{\lambda_2}^{n_2} \phi_{2}(x) \right), \label{u1} 
\end{align}
where the operators $f_{\lambda_i}^{n_i}, \, i=1,2$, which depend only on $\eta\cdot\partial_x+im\xi$, are defined as 
\begin{align}	
 f_{\lambda_i}^{n_i} = e^{\lambda_i(\eta\cdot\partial_x+im\xi)} (\eta\cdot\partial_x+im\xi)^{n_i}.	\label{s11}
\end{align}
Here $m$ is the continuous spin tachyon mass, $n_0, n_1$ and $n_2$ are non negative integers, $\lambda_1$ and $\lambda_2$ are two free real and dimensionful parameters and the operator $f_{\lambda_i}^{n_i}$ acts only on the first field in front of it. We require $n_0, n_1$ and $n_2$ to be non negative integers in order to be able to perform integration by parts in the cubic action. Using the ansatz (\ref{u1}) in (\ref{w2}) we find that $J_1$ is 
\begin{align}
	 &J_1(\eta,\xi,x) =  \frac{1}{2}n_0 (\eta\cdot\partial_x + im\xi)^{n_0-2} \times \nonumber \\
&	\left(
 (M_1^2+M_2^2+m^2)  f_{\lambda_1}^{n_1} \phi_{1}(x) \,\,\, f_{\lambda_2}^{n_2} \phi_{2}(x) - 2 f_{\lambda_1}^{n_1} \partial_x^\mu \phi_{1}(x) \,\,\, f_{\lambda_2}^{n_2} \partial_{x\mu} \phi_{2}(x) \right) \nonumber \\
& + \frac{1}{2} (\eta\cdot\partial_x + im\xi )^{n_0-1} \times \nonumber \\
&\left( n_1 (M_1^2+m^2) f_{\lambda_1}^{n_1-1} \phi_{1}(x) \,\,\, f_{\lambda_2}^{n_2} \phi_{2}(x) + n_2 (M_2^2+m^2)  f_{\lambda_1}^{n_1} \phi_{1}(x) \,\,\, f_{\lambda_2}^{n_2-1} \phi_{2}(x) \right. \nonumber \\ 
&\left. - (\lambda_1 f_{\lambda_1}^{n_1} + n_1 f_{\lambda_1}^{n_1-1} ) \partial_x^\mu \phi_{1}(x) \,\,\, f_{\lambda_2}^{n_2} \partial_{x\mu} \phi_{2}(x)  - f_{\lambda_1}^{n_1} \partial_x^\mu \phi_{1}(x)\,\,\, ( \lambda_2 f_{\lambda_2}^{n_2} + n_2 f_{\lambda_2}^{n_2-1}) \partial_{x\mu} \phi_{2}(x) \right), \label{s10} 
\end{align}
and 
\begin{align}
	\rho = \lambda_1(M_1^2+m^2) + \lambda_2(M_2^2+m^2). \label{s12}
\end{align}
Requiring $J_1$ to be local gives rise to two cases, either $n_0 = n_1 = n_2 = 0$ and $\lambda_1+\lambda_2=0$ or $n_0\ge 2$ and $n_1,n_2\ge 0$. In any case we must also require a non vanishing $\rho$ so that 
\begin{align}
	\frac{\lambda_1}{\lambda_2} \not= - \frac{M_2^2+m^2}{M_1^2+m^2}. \label{u2}
\end{align}
We then have a cubic vertex depending on three integer parameters $n_0, n_1$ and $n_2$ and one dimensionful parameter $\lambda_1$ or $\lambda_2$ since we can solve (\ref{s12}) for one of them. In the first case we get a very simple vertex. Calling  $\lambda_1=-\lambda_2\equiv\lambda$ we find   
\begin{align}
	J(\eta,\xi,x) =  e^{\lambda(\eta\cdot\partial_x+im\xi)} \phi_1(x) \,\,\,e^{-\lambda(\eta\cdot\partial_x+im\xi)} \phi_2(x),  \label{u3} 
\end{align}
and $\rho = \lambda(M_1^2-M_2^2)$ so that the masses of the scalar fields must be different. 


We can now take the continuous spin particle limit in (\ref{s3})-(\ref{w2}), (\ref{u1})-(\ref{s12}) by removing the integral in $\xi$ and setting $m=\xi=0$. We also take the scalar fields $\phi_i$ and the current $J$ to be real. The cubic vertex is now
\begin{align}
	 S_c = g \int d\eta \,dx \,\, \delta^{\prime}(\eta^2+\mu^2) \, \Psi(\eta,x) \, J(\eta,x), \label{u6}
\end{align}
and the currents are
\begin{align}
	&J_0(\eta,x) = (\eta\cdot\partial_x)^{n_0} \left( f_{\lambda_1}^{n_1} \phi_{1}(x) \,\,\, f_{\lambda_2}^{n_2} \phi_{2}(x) \right), \label{u9}\\
	 &J_1(\eta,x) =\frac{1}{2}n_0 (\eta\cdot\partial_x )^{n_0-2} \left(
 (M_1^2+M_2^2)  f_{\lambda_1}^{n_1} \phi_{1}(x) \,\,\, f_{\lambda_2}^{n_2} \phi_{2}(x) - 2 f_{\lambda_1}^{n_1} \partial_x^\mu \phi_{1}(x) \,\,\, f_{\lambda_2}^{n_2} \partial_{x\mu} \phi_{2}(x) \right) \nonumber \\
& + \frac{1}{2} (\eta\cdot\partial_x)^{n_0-1} \left( n_1 M_1^2 f_{\lambda_1}^{n_1-1} \phi_{1}(x) \,\,\, f_{\lambda_2}^{n_2} \phi_{2}(x) + n_2 M_2^2  f_{\lambda_1}^{n_1} \phi_{1}(x) \,\,\, f_{\lambda_2}^{n_2-1} \phi_{2}(x) \right. \nonumber \\ 
&\left. - (\lambda_1 f_{\lambda_1}^{n_1} + n_1 f_{\lambda_1}^{n_1-1} ) \partial_x^\mu \phi_{1}(x) \,\,\, f_{\lambda_2}^{n_2} \partial_{x\mu} \phi_{2}(x) - f_{\lambda_1}^{n_1} \partial_x^\mu \phi_{1}(x) \,\,\, ( \lambda_2 f_{\lambda_2}^{n_2} + n_2 f_{\lambda_2}^{n_2-1}) \partial_{x\mu} \phi_{2}(x)  \right),
	\label{u7} 
\end{align}
where now 
\begin{align}
	f_{\lambda_i}^{n_i} = e^{\lambda_i \eta\cdot\partial_x} (\eta\cdot\partial_x)^{n_i}. 
\end{align}
Then (\ref{s12}) becomes
\begin{align}
	\rho = \lambda_1 M_1^2 + \lambda_2 M_2^2, \label{s16}
\end{align}
so that $\lambda_1/\lambda_2 \not= - M_2^2/M_1^2$. 

For the case $n_0=n_1=n_2=0$, calling  $\lambda_1=-\lambda_2\equiv\lambda$, we have 
\begin{align}
	J(\eta,x) =  e^{\lambda\eta\cdot\partial_x} \phi_1(x) \,\,\,e^{-\lambda\eta\cdot\partial_x} \phi_2(x),  \label{u10} 
\end{align}
and $\rho = \lambda(M_1^2-M_2^2)$ so that the masses of the scalar fields must be different. This vertex has no free parameters. This is precisely the result found in \cite{Bekaert:2017xin} using BBvD-like currents \cite{Bekaert:2009ud}. 

The simplest solution for massive fields with the same mass $M$ has $n_0=2, n_1=n_2=0$ and the current is 
\begin{align}
	&J(\eta,x) =   (\eta\cdot\partial_x)^2 \left(e^{\lambda_1\eta\cdot\partial_x} \phi_{1}(x) \,\, e^{\lambda_2\eta\cdot\partial_x} \phi_{2}(x) \right)  \nonumber\\
	&  	+ (\eta^2+\mu^2) \left[ 2M^2  e^{\lambda_1\eta\cdot\partial_x} \phi_{1}(x) \,\,\,  e^{\lambda_2\eta\cdot\partial_x} \phi_{2}(x) - 2 e^{\lambda_1 \eta\cdot\partial_x} \partial_x^\mu \phi_{1}(x) \,\,\, e^{\lambda_2 \eta\cdot\partial_x} \partial_{x\mu} \phi_{2}(x) \right. \nonumber\\
	&\left. 
	- \frac{1}{2} \eta\cdot\partial_x \left( (\lambda_1+\lambda_2)  e^{\lambda_1\eta\cdot\partial_x} \partial_x^\mu \phi_{1}(x) \,\,\, e^{\lambda_2 \eta\cdot\partial_x} \partial_{x\mu} \phi_{2}(x)               \right) \right], \label{s18} 
\end{align}
and $\rho=(\lambda_1+\lambda_2)M^2$ so that $\lambda_1\not= -\lambda_2$ and $M\not= 0$. For scalar fields this is the situation analysed in \cite{Metsaev:2017cuz} using the oscillator formalism in the light-cone gauge. We can also consider vertices which are symmetric or antisymmetric by the interchange of $\phi_{1}$ and $\phi_{2}$. If we now take the limit $\rho=0$, which implies that $\lambda_1=-\lambda_2\equiv\lambda$, the continuous spin particle turns into an infinite tower of massless particles each one appearing once in $\Psi(\eta,x)$. The currents which are symmetric and anti-symmetric by the interchange of $\phi_{1}$ and $\phi_{2}$ are then 
\begin{align}
	&J_\pm (\eta,x) = (\eta\cdot\partial_x)^2 \left( e^{\lambda\eta\cdot\partial_x} \phi_{1}(x) \,\, e^{-\lambda\eta\cdot\partial_x} \phi_{2}(x)  \pm e^{\lambda\eta\cdot\partial_x} \phi_{2}(x) \,\, e^{-\lambda\eta\cdot\partial_x} \phi_{1}(x) \right)  \nonumber\\
	&  	+ (\eta^2+\mu^2) \left( 2M^2  e^{\lambda\eta\cdot\partial_x} \phi_{1}(x) \,\,\,  e^{-\lambda\eta\cdot\partial_x} \phi_{2}(x) - 2 e^{\lambda \eta\cdot\partial_x} \partial_x^\mu \phi_{1}(x) \,\,\, e^{-\lambda \eta\cdot\partial_x} \partial_{x\mu} \phi_{2}(x)  \right. \nonumber\\
	& \left. \pm 2M^2  e^{\lambda\eta\cdot\partial_x} \phi_{2}(x) \,\,\,  e^{-\lambda\eta\cdot\partial_x} \phi_{1}(x) \mp 2 e^{\lambda \eta\cdot\partial_x} \partial_x^\mu \phi_{2}(x) \,\,\, e^{-\lambda \eta\cdot\partial_x} \partial_{x\mu} \phi_{1}(x)  \right).	\label{s21}
\end{align}
If we expand the exponentials we find that $J_+$, which is even by the interchange of $\lambda$ and $-\lambda$, has only even powers of $\eta^\mu$. In the cubic action the integral over $\eta^\mu$ will select only even powers in $\Psi(\eta,x)$ so that $J_+$ couples to even spins in $\Psi(\eta,x)$. A similar reasoning for $J_-$ shows that it couples only to odd spins in $\Psi(\eta,x)$. This agrees with the results presented in  \cite{Metsaev:2017cuz} for a continuous spin particle and two massive scalar fields with the same mass when $\rho\rightarrow 0$. 

\section{Conclusions \label{l10}}


We have presented an action for continuous spin tachyonic gauge fields. The analysis of its physical contents agree with that obtained in the oscillator formalism \cite{Metsaev:2016lhs}. There the starting point is a collection of totally symmetric tensor fields which are double-traceless and are contracted with creation and annihilation operators. Here we start with a complex field $\Psi(\eta,\xi,x)$ living on a cotangent bundle  which can be expanded in terms of totally symmetric spacetime fields which are unconstrained. Since we get the same results we could hope to show that both formulations are completely equivalent. In the higher spin case \cite{Metsaev:2011iz} it was found a correspondence between terms in both actions \cite{Rivelles:2014fsa} after the integrals over $\eta^\mu$ were performed. However no direct map between the gauge field and the oscillators  was found so it is unlikely that such map exists in the continuous spin case. 

We have presented an analysis of cubic vertices for one continuous spin tachyon and two massive scalar particles and for one continuous spin particle and two massive scalar particles. In both cases we found that in general they depend on one dimensionful parameter and three non negative integer parameters. It would be interesting to find a systematic way to analyze these vertices. 

We have shown that there is now an improved understanding of the local symmetries on the cotangent bundle which allowed us to find much more suitable gauge fixing conditions not only in the tachyonic case but also for the continuous spin particle. This will now hopefully allow us to extend the continuous spin particle results to AdS spaces. Also, the extension of the the analysis presented in this paper to the fermionic  case along the lines of \cite{Najafizadeh:2015uxa} seems feasible.

Finally it should be remarked that a full analysis of cubic vertices has to be performed. It seems doable in the present context and it will improve our understanding of how continuous spin tachyons and continuous spin particles interact with other fields and among themselves.

\acknowledgments

I would like to thank R. Metsaev for very useful discussions about continuous spin particle vertices and M. Najafizadeh for pointing out a wrong sign for the quartic Casimir operator. 
I also would like to thank F.M.S. Freitas for checking some of the calculations of this paper.  This work was supported by FAPESP Grants 2014/18634-9 and 2019/21281-4. 


\end{document}